\documentclass[%
 aip,
 jmp,%
 amsmath,amssymb,
 reprint,%
]{revtex4-1}

\usepackage{graphicx}
\usepackage{hyperref}
\pdfoutput=1
\usepackage[all]{hypcap}
\usepackage{xcolor}
\hypersetup{
    colorlinks,
    linkcolor={red!50!blue},
    citecolor={blue!50!blue},
    urlcolor={blue!80!blue}
}
\usepackage{dcolumn}
\usepackage{bm}

\begin{document}

\preprint{AIP/123-QED}

\title {Electrical excitation of silicon-vacancy centers in single crystal
diamond}

\author{Amanuel M. Berhane}
\author{Sumin Choi}
\affiliation{ 
School of Physics and Advanced Materials, University of Technology Sydney, Ultimo, New South Wales 2007, Australia
}%


\author{Hiromitsu Kato}
\author{Toshiharu Makino}
\affiliation{Energy Technology Research Institute, AIST, Tsukuba, Ibaraki 305-8568, Japan}

\author{Norikazu Mizuochi}
\affiliation{Graduate School of Engineering Science, Osaka University 1-3, Machikane-yama, Toyonaka-city, Osaka, 560-8531, Japan}

\author{Satoshi Yamasaki}
\affiliation{Energy Technology Research Institute, AIST, Tsukuba, Ibaraki 305-8568, Japan}

\author{Igor Aharonovich}%
 \email{igor.aharonovich@uts.edu.au}
\affiliation{ 
School of Physics and Advanced Materials, University of Technology Sydney, Ultimo, New South Wales 2007, Australia
}%

\date{\today}

\begin{abstract}
Electrically driven emission from negatively charged silicon-vacancy (SiV)$^{-}$ centers in single crystal diamond is demonstrated. The SiV centers were generated using ion implantation into an i region of a p-i-n single crystal diamond diode. Both electroluminescence and the photoluminescence signals exhibit the typical emission that is attributed to the (SiV)$^{-}$ centers. Under forward and reversed biased PL measurements, no signal from the neutral (SiV)$^{0}$ defect could be observed. The realization of electrically driven (SiV)$^{-}$ emission is promising for scalable nanophotonics devices employing color centers in single crystal diamond. 
\end{abstract}

\maketitle

Color centers in diamond are promising building blocks for a variety of applications, including quantum information processing and integrated nanophotonics \cite{Aharonovich2014,Childress2013,Becher2014,Hausmann2012}. Recently, a major research focus has been dedicated to identify bright single photon emitters with narrow line widths and fast excited state lifetimes at the near infrared spectral range. One emerging candidate is the negatively charged silicon vacancy (SiV)$^{-}$ color center that consists of an interstitial silicon splitting two vacancies in the diamond lattice \cite{Goss1996,Neu2011,Matthiesen2012,Sipahigil2014}. The SiV exhibits photostable room temperature operation with narrow zero phonon line (ZPL) ($\sim$ 1-3 nm), fast excited state lifetime ($\sim$ 1 ns), polarized excitation and optical readout of its spin state at low temperature\cite{Muller2014,Neu2012}. These excellent photophysical properties are therefore attractive for applications spanning bio-imaging\cite{Merson2013}, nanophotonics and quantum information processing\cite{Childress2013,OBrien2010}.

Although the SiV centers have been studied in details under optical excitation, electroluminescence properties of this center in a single crystal diamond remain unexplored. Electrical excitation is important since it opens pathways to engineer scalable devices employing electrically driven emitters and realize efficient packaging on a single chip. Furthermore, it enables studying charge injection and dynamics between various charge states of a particular defect \cite{Ellis2011,Stevenson2012,Mizuochi2012,Doi2014}. In this work we report on electrical excitation of engineered SiV defects in a single crystal diamond. In particular, we show that the same emission is obtained using optical and electrical excitation - a unique feature that has not been demonstrated in single crystalline diamond yet \cite{Liang2007a}.

\begin{figure}[h]
\begin{center}
\includegraphics[scale=1]{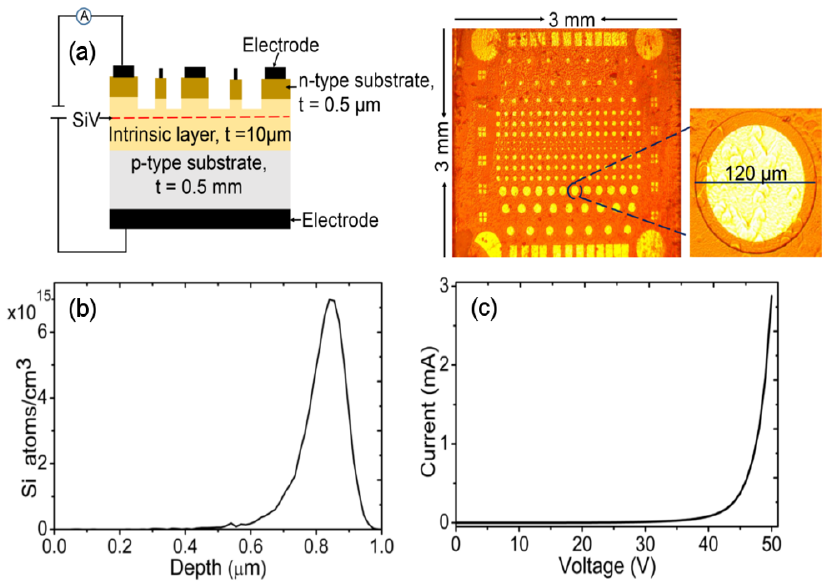}
\caption{(a) Schematic illustration of a single crystal PIN diode with implanted Si atoms and an optical image of the device. The diameters of the n-type diamond mesa is $120 \mu m$ and the metallic contacts on top  is $100 \mu m$  (b) Monte Carlo depth profile of ion implanted $Si$ atoms into diamond  obtained using SRIM calculations. The end of range is estimated around 820$nm$. (c) I-V-characteristic plot showing diode rectification at a forward threshold voltage of 43$V$ at room temperature}
\label{fig:introduction}
\end{center}
\end{figure}

The starting material is a single crystal diamond PIN diode \cite{Mizuochi2012,Kato2013}. The diode structure consists of a p-type diamond substrate layer of thickness 0.5 $mm$, intrinsic diamond layer of 10 $\mu m$ and 0.5 $\mu m$ n-type diamond mesas. Detailed growth conditions were described elsewhere\cite{Mizuochi2012}. As a contact electrode, 30 nm $Ti$/ 100 nm $Pt $/ 200 nm $Au$ were deposited on both the n- and p-type sides of the device. Fig.  \ref{fig:introduction}(a) shows the schematic of the device (left) and the optical microscope image of a single 120 $\mu m$ diameter circular mesa (right). The dark circle around the n-type layer is a two-dimensional image depth resolution depiction of the  pillar height. 

\smallskip

To introduce the silicon atoms into the diamond, ion implantation is employed. Monte Carlo simulation is used to estimate the implantation energy so that the silicon atoms will predominantly reside within the i-region of the device. Fig. \ref{fig:introduction}(b) shows the Stopping and Range of Ions in Matter (SRIM) simulation under 1.5 $MeV$ acceleration energy, which results in approximately 820 nm end of range. A small dose of 1$\times$ 10$^{11}$ $\frac{atoms}{cm^{2}}$ is implanted over the whole area of the sample. The diode property of the device is then characterized via I-V rectification curve under forward bias and the result is shown in fig. \ref{fig:introduction}(c). The threshold forward current of 0.89 $mA$ at room temperature corresponds to current density of 7.9 $\frac{A}{cm^{2}}$ at the specific device showing rectification at threshold voltage of 43 $V$. This high threshold voltage driving a relatively low current density into the device is  attributed to the high specific contact resistance between phosphorous doped n-type diamond mesas and top side contacts \cite{Kato2008}. 

Elctroluminescence (EL) from the fabricated SiV centers is carried out using external voltage source to inject carriers and scanning confocal microscope with a high numerical aperture objective (0.7, Nikon, $\times$100). The collected light is  passed through a dichroic mirror and focused onto a graded index fiber with a core size of 62.5 $\mu$m that acts as a confocal aperture. The collected emission  splits into an avalanche photo detector and a spectrometer. Photoluminescence (PL) measurements were carried out on the same set-up using a continuous wave 532 nm excitation laser.  

\smallskip
\begin{figure}[h]
\begin{center}
\includegraphics[scale=1]{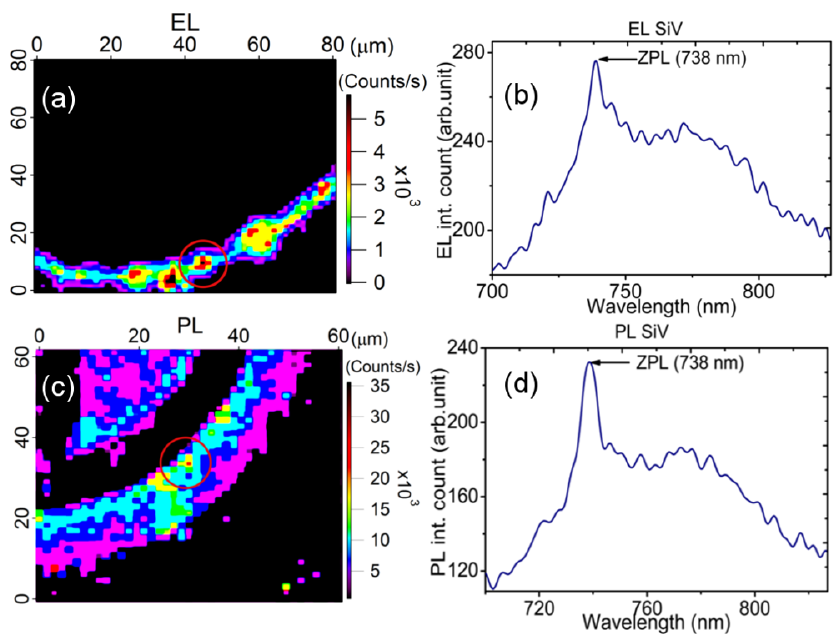}
\caption{ (a) Electroluminsce map of a 80$\mu m$$^{2}$ area showing luminescence from the edge of pillar (b) EL spectrum from the circled bright spot of the EL map. The EL spectrum is collected at forward bias of 50 $V$ used to inject current of 2.9 $mA$ into the device (c) Photoluminscence map of a 60$\mu m$$^{2}$ area, exhibiting comparable emission from around the edge of the pillar. (d) PL spectrum from the circled bright spot of the PL map. The excitation is performed using a  532 $nm$ cw laser at 867 $\mu W$} 
\label{fig:el}
\end{center}
\end{figure}

Fig. \ref{fig:el}(a) and \ref{fig:el}(b) show the EL confocal map as well as corresponding spectra from bright centers identified at the edges of the mesas under a current of 2.9 mA. The EL map indicates the recombination zone with lower carrier concentration at equilibrium is the i-layer of the device with bright circular luminescence observed around the pillar. This is in agreement with EL studies from ion implanted NV centers in diamond as reported in Ref \cite{Mizuochi2012}. The EL spectrum in fig. \ref{fig:el}(b) corresponds to bright spot circled on the map as shown in fig. \ref{fig:el}(a). This spectrum shows a ZPL at 738 $nm$ which corresponds to the ZPL of the (SiV)$^{-}$ center in diamond. 

Fig. \ref{fig:el}(c) and \ref{fig:el}(d) show the confocal PL map and corresponding spectra of the devices under optical excitation using a 532 nm laser at 867 $\mu$W. Like in the case of the EL measurement, bright spots are observed at the edges of the mesa. The measured PL spectrum exhibits a similar peak at 738 nm. This is a direct evidence that the (SiV)$^{-}$ color center is excited both optically and electrically in a single crystal diamond. The observation of the same center with the same charge state both optically and electrically in diamond is important for practical device engineering and is an advantageous feature of the (SiV)$^{-}$ defect. Indeed, while NV centers were excited electrically, only the NV$^{0}$ charge state is visible under electrical excitation while the NV$^{-}$ defect can be triggered optically \cite{Mizuochi2012,Kato2008,Himics2014,Lohrmann2011} .

\smallskip
\begin{figure}[h]
\begin{center}
\includegraphics[scale=1]{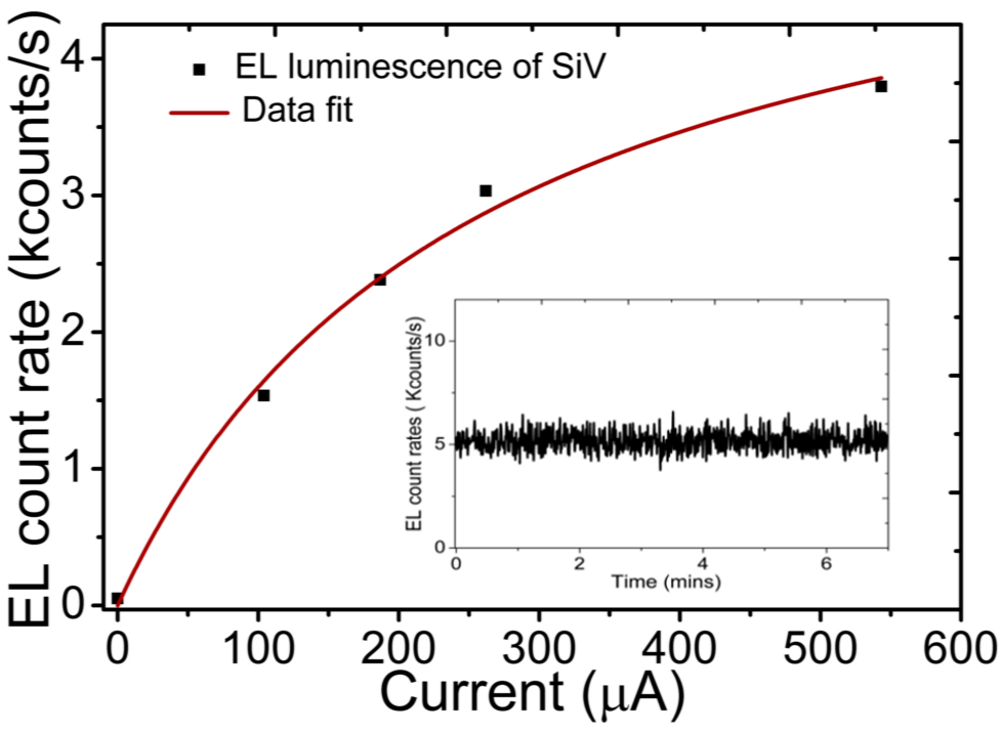}
\caption{Electrical driven luminescence saturation  measurement of the stable SiV vacancy together with fitting curve indicated by the solid red line. Fitting results yield highest power output in terms of EL count rates as 5669 $counts/s$ where the injection current of 0.3 $mA$. Inset shows electrical stability measurement of the negative charge state of SiV color center showing stable emission for more than 6 min}
\label{fig:power}
\end{center}
\end{figure}

Fig. \ref{fig:power} shows the saturation measurement of (SiV)$^{-}$ plotted as a function of applied current. The fitting equation is given by $r=r_{sat}\frac{I}{I_{sat}+I}$ where $r$ is arbitrary EL count rate for injection current ($I$), $r_{sat}$ is the maximum electrically driven luminescence count rate that can be attained in this device at saturation current of $I_{sat}$. Fitting results suggest that $r_{sat}$ is ~ 5.7 $kcounts/s$ for $I_{sat}$ of approximately 0.3 $mA$. While the emission intensity is not strong, the values are comparable with emission recorded optically from ion implanted (SiV)$^{-}$ defects \cite{Wang2005,Tamura2014,Pezzagna2011a}.

Electrical stability of the (SiV)$^{-}$ center under forward bias is also studied, to ensure a constant emission of photons is generated. The inset of Fig. 3 shows that EL count rates as a function of time for fixed injection of carriers. The emitters remain stable for the duration of the measurement (minutes), with no blinking or bleaching being observed. The stable emission is promising for utilizing the (SiV)$^{-}$ for quantum information and nanophotonics applications.

\smallskip
\begin{figure}[h]
\begin{center}
\includegraphics[scale=0.3]{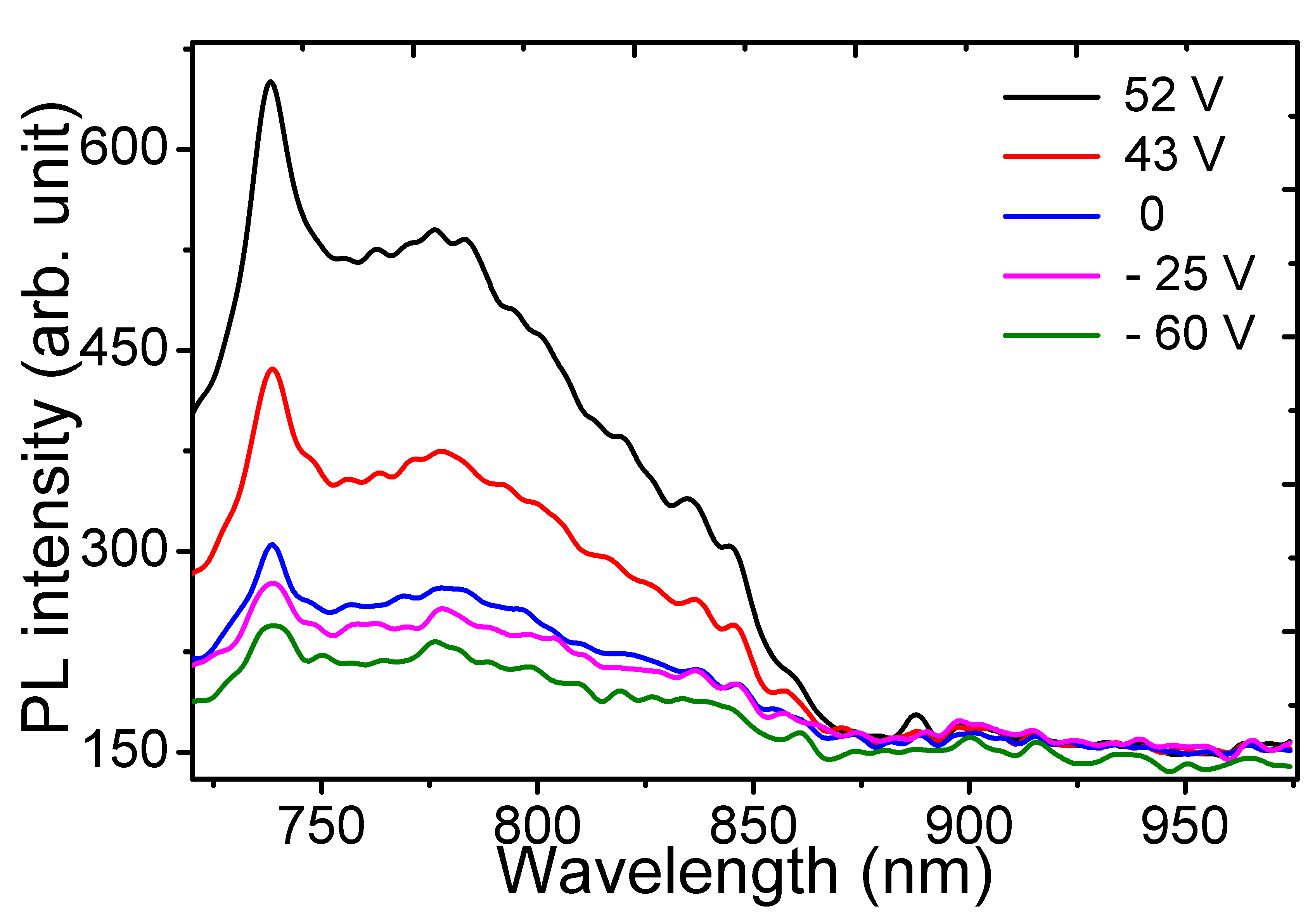}
\caption{Forward and reverse biased PL measurement show reduction in the intensity of the (SiV)$^{-}$ ZPL. The bias voltage is changed from 52V to -100 V varying the corresponding forward and reverse current in the device. Inset shows the normalized plot  where intensity from (SiV){$^{-}$} is kept to one so changes in regions around 946 nm could be studied. As shown, no obvious emission is observed in the region of interest.}
\label{fig:frd}
\end{center}
\end{figure}

Finally, we studied the behavior of the SiV defect under forward and reverse biased PL measurements. These studies were fruitful to understand the switching dynamics of the NV defect in diamond \cite{Doi2014, Kato2013}. For the case of the NV center, applying negative bias under constant green excitation resulted in switching the NV state from the negatively charged to neutral state. Fig. \ref{fig:frd} shows intensity varying emission from SiV centers as a result of forward and reverse biasing under 532 nm excitation laser. The bias voltage is varied between 52 V to -100 V injecting different amount of  forward and reverse current to the sample. As expected, under forward bias PL, the intensity of the (SiV)$^{-}$ ZPL decreases with decreased bias voltage. Under an increased reversed bias, the (SiV)$^{-}$ ZPL intensity further decreases - likely due to injection of holes into the defect. As shown, however, no clear evidence of (SiV)$^{0}$ emission at $\sim$ 946 nm was observed under these conditions \cite{DHaenens-Johansson2011}. This can be explained by the relatively low quantum efficiency of this defect or the low electron absorption cross section of the defect. 

To summarize, electrical excitation of the (SiV)$^{-}$ in a single crystal diamond is demonstrated. Both EL and PL excited the (SiV)$^{-}$ defect, which is highly important for scalable nanophotonics applications. The electrically driven emission is stable at room temperature and can be detected under current density injection of 7.9 ${A}/{cm^{2}}$. Forward and reverse biased PL measurement combined with short-circuited PL measurements have provided insight for the excitation mechanisms of the (SiV)$^{-}$ color centers, However, the (SiV)$^{0}$ or switching between the two charge states was not observed. Further work is therefore required to understand the dynamics of this defect under electrical excitation. In addition, investigation of ways to improve the creation efficiency of the SiV emitters using ion implantation is also required to achieve electrical excitation of a single SiV center.  

The authors thank Milos Toth for useful discussions, Brett Johnson for the assistance with ion implantation and The Department of Electronic Materials Engineering at the Australian National University for the access to the implantation facilities. Igor Aharonovich is the recipient of an Australian Research Council Discovery Early Career Research Award (Project number DE130100592).




\bibliography{turial} 
\end{document}